\documentclass[aps,prd,showpacs,amsmath,amssymb,twocolumn,superscriptaddress,preprintnumbers,nofootinbib]{revtex4-1}
\pdfoutput=1
\usepackage{color,epsfig}
\usepackage{amsthm}
\usepackage{amsmath}
\usepackage{mathrsfs}
\usepackage{graphicx,subfigure}
\usepackage{dcolumn}
\usepackage{bm}
\usepackage{slashed}
\usepackage{calligra}
\usepackage{hyperref}
\usepackage{tabularx,colortbl}

\definecolor{nicered}{rgb}{0.5,0.,0.}
\definecolor{nicegreen}{rgb}{0.,0.5,0.}
\definecolor{niceblue}{rgb}{0.,0.,0.5}
\hypersetup{colorlinks,citecolor=nicegreen,linkcolor=nicered,urlcolor=niceblue}

\DeclareMathAlphabet{\mathcalligra}{T1}{calligra}{m}{n}
\DeclareFontShape{T1}{calligra}{m}{n}{<->s*[2.5]callig15}{}

\allowdisplaybreaks

\newcommand{\be}{\begin{eqnarray}}
\newcommand{\ee}{\end{eqnarray}}
\newcommand{\bea}{\begin{eqnarray}}
\newcommand{\eea}{\end{eqnarray}}

\newcommand{\GeV}{{~\rm GeV}}

\newcommand{\TeV}{{~\rm TeV}}

\newcommand{\zprime}{Z^\prime}
\newcommand{\mzp}{m_{\zprime}}
\newcommand{\vprime}{v_\Phi}
\newcommand{\gp}{g^\prime}

\begin{document}

\title{Predictions for Lepton Flavor Universality Violation in Rare B Decays \texorpdfstring{\\}{} in Models with Gauged \texorpdfstring{\boldmath $L_\mu - L_\tau$}{Lmu - Ltau}}

\author{Wolfgang Altmannshofer}
\email{waltmannshofer@perimeterinstitute.ca}
\affiliation{Perimeter Institute for Theoretical Physics 31 Caroline St. N, Waterloo, Ontario, Canada N2L 2Y5.}
\author{Itay Yavin}
\email{iyavin@perimeterinstitute.ca}
\affiliation{Perimeter Institute for Theoretical Physics 31 Caroline St. N, Waterloo, Ontario, Canada N2L 2Y5.}
\affiliation{Department of Physics \& Astronomy, McMaster University 1280 Main St. W. Hamilton, Ontario, Canada, L8S 4L8.}

\begin{abstract}
A recent proposal for explaining discrepancies in angular observables in the rare decay $B\to K^* \mu^+\mu^-$ with a gauged $L_\mu - L_\tau$ current carried with it the prediction of lepton flavor universality violation in related $B$-meson decays. This prediction gained empirical support with a subsequent hint for lepton flavor universality violation in the $B\to K \ell^+\ell^-$ decay by LHCb. In this short paper we fully quantify the prediction including the associated uncertainties. We also provide new predictions for a variety of additional observables sensitive to lepton flavor universality violation in $B$-meson decays. 
\end{abstract}


\maketitle

\section{Introduction} \label{sec:intro}

In the past two years evidence has accumulated pointing to discrepancies between theoretical predictions and experimental results in angular observables in the rare decay $B\to K^* \mu^+\mu^-$~\cite{Aaij:2013qta,LHCb-CONF-2015-002}. It was noted early on~\cite{Descotes-Genon:2013wba,Altmannshofer:2013foa,Beaujean:2013soa,Hurth:2013ssa} that new physics (NP) contributing to the operator, 
\begin{equation}
\label{eqn:Hamiltonian}
 \mathscr{H}_\text{eff} = \Delta \mathscr{C}_9^\mu (\bar s \gamma_\alpha P_L b)(\bar \mu \gamma^\alpha \mu)~,
\end{equation}
can explain the discrepancies. This operator consists of a left-handed quark current and a vector current for the muons.
The theoretical predictions of the Standard Model (SM) involve non-perturbative QCD effects that could mimic a NP contribution to exactly such an operator. It is not easy to reliably estimate some parts of the non-perturbative QCD effects~\cite{Khodjamirian:2010vf,Jager:2012uw,Lyon:2014hpa,Descotes-Genon:2014uoa,Jager:2014rwa}. Thus, the aforementioned discrepancies might still be the result of underestimated SM contributions. Nevertheless, a variety of NP models have been proposed in the literature as the origin of the operator in Eq.~(\ref{eqn:Hamiltonian})~\cite{Descotes-Genon:2013wba,Altmannshofer:2013foa,Gauld:2013qba,Buras:2013qja,Gauld:2013qja,Datta:2013kja,Buras:2013dea,Altmannshofer:2014cfa,Biancofiore:2014wpa,Buras:2014fpa,Altmannshofer:2014rta,Gripaios:2014tna,Crivellin:2015mga,Crivellin:2015lwa,Sierra:2015fma,Celis:2015ara,Belanger:2015nma,Niehoff:2015iaa}.

In Ref.~\cite{Altmannshofer:2014cfa} we presented a model with an extra vector-boson associated with the gauging (and spontaneous breaking) of muon-number minus tau-number, $L_\mu - L_\tau$. Gauging of $L_\mu - L_\tau$ guarantees a vectorial coupling of the vector-boson to muons. The new vector-boson can couple to the different quark flavours indirectly through its coupling to very heavy colored fermions that mix with the quarks as in Ref.~\cite{Fox:2011qd}. If the mass of the new vector-boson is sufficiently heavy compared to the $B$-meson mass, then it can be integrated out of the low-energy theory to yield the operator in Eq.~(\ref{eqn:Hamiltonian}). 

One clean prediction of this model, first made in Ref.~\cite{Altmannshofer:2014cfa}, is a $\sim 20\%$ suppression of the rate of the inclusive $B \to X_s \mu^+\mu^-$ decay relative to the electron mode $B \to X_s e^+e^-$ which remains SM-like. The structure of $L_\mu - L_\tau$ also predicts a corresponding $\sim 20\%$ enhancement of the decay rate of the tauonic mode with respect to its SM prediction. The predicted suppression in the muonic mode has subsequently received experimental support with the recent hints for violation of lepton flavor universality (LFU) in the exclusive decay $B \to K \ell^+\ell^-$ by LHCb~\cite{Aaij:2014ora}
\be \label{eq:RK}
 R_K &=& \frac{\text{BR}(B \to K \mu^+\mu^-)}{\text{BR}(B \to K e^+e^-)}  \\\nonumber &=& 0.745^{+0.090}_{-0.074} ~\text{(stat)}~\pm 0.036 ~\text{(syst)} ~,
\ee
where the branching ratios refer to a dimuon invariant mass squared, $q^2$, between 1~GeV$^2 < q^2 < 6$~GeV$^2$. 
A number of analyses have appeared since, interpreting the hint of lepton flavor universality violation~(\ref{eq:RK}) in the context of new physics models~\cite{Alonso:2014csa,Hiller:2014yaa,Ghosh:2014awa,Biswas:2014gga,Hurth:2014vma,Glashow:2014iga,Hiller:2014ula,Bhattacharya:2014wla,Niehoff:2015bfa,Becirevic:2015asa,Alonso:2015sja,Greljo:2015mma,Calibbi:2015kma}.

In this short paper we wish to sharpen the prediction for lepton flavor universality violation made in Ref.~\cite{Altmannshofer:2014cfa}, including the theoretical uncertainties associated with it. Furthermore, we provide a variety of new predictions for lepton flavour universality violation in other, independent, observables.

\section{The Model} \label{sec:model}

We briefly review the defining components of the model and refer the reader to Ref.~\cite{Altmannshofer:2014cfa} for more details. We extend the SM by a new $U(1)^\prime$ gauge group with a new gauge-boson, $\zprime$, corresponding to muon-number minus-tau number, $L_\mu - L_\tau$. The gauge group is anomaly free with the SM particle content. The $L_\mu - L_\tau$ gauge symmetry is spontaneously broken by the vacuum expectation value $\vprime$ of a new heavy scalar $\Phi$ charged under $L_\mu - L_\tau$. The gauge-boson obtains a mass $\mzp = \gp \vprime$, with the $U(1)^\prime$ gauge coupling $\gp$. As we will see below, the $B$-meson anomalies are not very sensitive to either of these parameters, $\mzp$, $\gp$, or $\vprime$. Other observations~\cite{Altmannshofer:2014cfa, Altmannshofer:2014pba} constrain the model, but leave ample room in the parameter space defined by $\mzp$ and $\gp$ with $O(10) \GeV \lesssim \mzp \lesssim \text{few} \TeV$ with a corresponding increasing range in the allowed gauge 
coupling $\gp$. 

\begin{figure}
\includegraphics[scale=0.65]{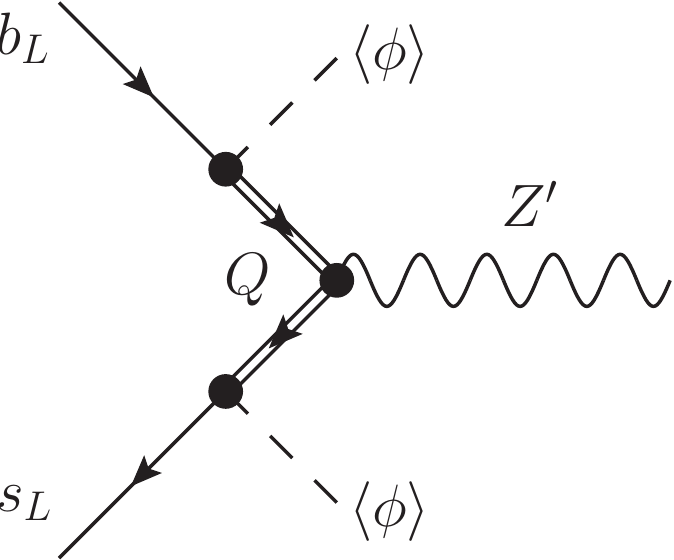}
\caption{Example diagram in the high energy theory that leads to flavor-changing effective couplings of the $Z^\prime$ to left-handed SM quarks. Similar diagrams involving other heavy colored fermions can lead to an effective coupling of the $Z^\prime$ to right-handed SM quarks.}
\label{fig:effective_coupling}
\end{figure}

The coupling of the $\zprime$ to quarks at low-energies is derived from an anomaly-free theory involving new heavy colored fermions, which are vector-like charged under the $U(1)^\prime$ and the SM gauge group. These heavy colored fermions couple to the SM quarks through Yukawa couplings with the scalar $\Phi$. The spontaneous breaking of the $U(1)^\prime$ leads to mass mixing between the SM quarks and the heavy fermions. In turn, the mass mixing results in an effective coupling of the $\zprime$ to SM quarks, as illustrated in Fig.~\ref{fig:effective_coupling}. The low-energy effective coupling is therefore determined by the type of heavy fermions present (electroweak singlets vs.\ doublets), their masses, and their Yukawa couplings to the SM quarks and $\Phi$. At the even lower energies scales relevant for $B$-meson decays, after the $Z^\prime$ has been integrated out, we are left with contributions to the Wilson coefficients of the semileptonic $b \to s \ell\ell$ operators with leptonic vector currents
\begin{equation} \label{eq:Heff}
\mathscr{H}_\text{eff} = \sum_{\ell = e, \mu, \tau} \Big[ \Delta \mathscr{C}_9^\ell (\bar s \gamma_\alpha P_L b) +  \Delta \mathscr{C}_9^{\prime \ell} (\bar s \gamma_\alpha P_R b) \Big] (\bar \ell \gamma^\alpha \ell) ~,
\end{equation}
with
\begin{eqnarray}
\label{eqn:def_C_9}
\Delta \mathscr{C}_9^e = 0 ~&,&~~ \Delta \mathscr{C}_9^{\mu} = - \Delta \mathscr{C}_9^{\tau} = \frac{Y_{Qb} Y_{Qs}^*}{2m_Q^2}~, \\
\label{eqn:def_Cp_9}
\Delta \mathscr{C}_9^{\prime e} = 0 ~&,&~~ \Delta \mathscr{C}_9^{\prime \mu} = - \Delta \mathscr{C}_9^{\prime \tau} = -\frac{Y_{Db} Y_{Ds}^*}{2m_D^2} ~,
\end{eqnarray}
where $m_Q$ and $m_D$ are the masses of heavy electroweak doublet quarks and heavy electroweak singlet down type quarks, respectively, and  $Y_{Qi}$, $Y_{Di}$ are their corresponding Yukawa couplings with $\Phi$, see~\cite{Altmannshofer:2014cfa}. In addition to the operator in~(\ref{eqn:Hamiltonian}), also its right-handed counterpart is generated in general. 
It is common to use dimensionless Wilson coefficients $\Delta C_9^{(\prime)}$ defined as
\begin{equation}
 - \frac{4G_F}{\sqrt{2}} V_{tb} V_{ts}^* \frac{e^2}{16\pi^2} \Delta C_9^{(\prime)} = \Delta \mathscr{C}_9^{(\prime)} ~.
\end{equation}
As alluded to earlier, interestingly the Wilson coefficients do not depend on $\mzp$, $\gp$, or $\vprime$, but only on the properties of the heavy fermions and their Yukawa couplings to SM quarks. This is so because the $\mzp^2$ that would  appear in the denominator from integrating out the $Z^\prime$, is exactly canceled by the product of $(\gp)^2$ (coming from each vertex) and $\vprime^2$ (coming from the mass mixing) in the numerator. In addition, the $L_\mu - L_\tau$ gauge symmetry forces $\Delta C_9^e = \Delta C_9^{\prime e}= 0$  and opposite sign between the muonic and tauonic contributions. We are only free to adjust $\Delta C_9^\mu$ and $\Delta C_9^{\prime \mu}$ in Eqs.~(\ref{eqn:def_C_9}) and~(\ref{eqn:def_Cp_9}) through the choice of Yukawa coupling and heavy fermion masses. This tight structure is the origin of the sharp prediction for the presence of LFU violations in $B$-meson decay: once $\Delta C_9^\mu$ and $\Delta C_9^{\prime \mu}$ are fixed to explain the angular anomalies in $B\to K^* \mu^+\
mu^-$, they immediately effect the branching ratio of $B\to K \mu^+\mu^-$ and various other observables in rare $B$-meson decays based on the $b \to s \mu\mu$ transition. All electron modes, $b \to s ee$, remain SM-like. In the following we will drop the lepton flavor superscript on the Wilson coefficients and denote the muon specific coefficients with $\Delta C_9$ and $\Delta C_9^\prime$.

\section{Probes of Lepton Flavor Universality}
\label{sec:LFU}

We consider observables in various decays based on the quark level $b \to s \ell\ell$ transition that are affected by the operators in Eq.~(\ref{eq:Heff}): the exclusive decays $B \to K \ell^+\ell^-$, $B \to K^* \ell^+\ell^-$, and $B_s \to \phi \ell^+\ell^-$, as well as the inclusive $B \to X_s \ell^+\ell^-$ decay.

\subsection{Exclusive Decays}
\label{sec:exc}

We consider the following ratios of branching ratios that are tests of lepton flavor universality~\cite{Hiller:2003js,Bobeth:2007dw,Hiller:2014ula}
\begin{eqnarray}
R_K &=& \frac{\text{BR}(B \to K \mu^+\mu^-)}{\text{BR}(B \to K e^+e^-)}  ~, \\
R_{K^*} &=& \frac{\text{BR}(B \to K^* \mu^+\mu^-)}{\text{BR}(B \to K^* e^+e^-)}  ~, \\
R_\phi &=& \frac{\text{BR}(B \to \phi \mu^+\mu^-)}{\text{BR}(B \to \phi e^+e^-)}  ~. 
\end{eqnarray}
In the SM, these ratios are expected to be unity to a very high accuracy.\footnote{Radiation of collinear photons off the leptons can lead to logarithmically enhanced QED corrections that can be different in the electron and muon modes. In the $R_K$ measurement performed by LHCb~\cite{Aaij:2014ora}, such effects are included in the event simulation.}
In the following, superscripts on the LFU observables indicate the bin of di-lepton invariant mass in which the observable is measured. For example, $R_K^{[1,6]}$ refers to the $q^2$ bin $1\,$GeV$^2 < q^2 < 6\,$GeV$^2$, etc.

In the $L_\mu - L_\tau$ framework, the above LFU ratios can be expressed in terms of the muon specific Wilson coefficients $\Delta C_9$ and $\Delta C_9^\prime$. For all three decays we consider one bin at low and high $q^2$ each. We provide explicit expressions in Appendix~\ref{sec:appendix1}. 

The ratios $R_\phi$ and $R_{K^*}$ depend on the left-handed and right-handed Wilson coefficients $\Delta C_9$ and $\Delta C_9^\prime$ in a very similar way. Differences only arise from $SU(3)$ flavor breaking and the rather sizable life time difference in the $B_s$-meson system. However, given the expected experimental sensitivities in the near future, one has to an excellent approximation $R_\phi \simeq R_{K^*}$. Comparing $R_\phi$ and $R_{K^*}$ with $R_K$ allows in principle to disentangle contributions from left-handed and right-handed quark currents~\cite{Hiller:2014ula}.

In addition to the branching ratio ratios, we also consider lepton flavor universality tests using angular observables in the $B \to K^* \ell^+\ell^-$ decay. As some angular observables have zero crossings we do not use lepton flavor ratios but rather construct lepton flavor differences
\begin{eqnarray}
D_{A_\text{FB}} &=& A_\text{FB}(B \to K^* \mu\mu) - A_\text{FB}(B \to K^* ee) ~, \\
D_{F_L} &=& F_L(B \to K^* \mu\mu) - F_L(B \to K^* ee) ~, \\
D_{S_5} &=& S_5(B \to K^* \mu\mu) - S_5(B \to K^* ee) ~.
\end{eqnarray}
Here, $A_\text{FB}$ is the lepton forward-backward asymmetry, $F_L$ is the longitudinal polarization fraction of the $K^*$ vector meson, and the angular observable $S_5$ is related to the ``$B \to K^* \mu^+\mu^-$ anomaly'' in the observable $P_5^\prime = S_5/\sqrt{F_L (1-F_L)}$~\cite{Aaij:2013qta}.
Other probes of lepton flavor universality have been considered in the recent literature. For instance: lepton flavor ratios of angular observables in $B \to K^* \ell^+ \ell^-$~\cite{Altmannshofer:2014rta,Jager:2014rwa} and lepton flavor specific shifts of the zero crossings of the angular observables $A_\text{FB}$ and $S_5$~\cite{Jager:2014rwa}.

In contrast to $B \to K^* \ell^+\ell^-$, the $B_s \to \phi \ell^+\ell^-$ decay is not self-tagging. Therefore, the observables $D_{A_\text{FB}}$ and $D_{S_5}$ cannot be easily accessed in the $B_s \to \phi \ell^+\ell^-$ decay.
We therefore concentrate on $B \to K^* \ell^+\ell^-$.

At high $q^2$, the angular observables $A_\text{FB}$, $F_L$, and $S_5$ do not differ significantly from their SM predictions in regions of NP parameter space that provide a good fit of the anomalies~\cite{Altmannshofer:2014rta}. Thus, we consider the corresponding LFU differences only in the low $q^2$ region. 
Expressions for the new $B \to K^* \ell^+\ell^-$ LFU differences in terms of the Wilson coefficients $\Delta C_9$ and $\Delta C_9^\prime$ are given in Appendix~\ref{sec:appendix1}.

The new observables provide additional means to test lepton flavor universality and, in the case that evidence for violation of lepton flavor universality can be established, can help to distinguish between left-handed and right-handed currents.

\subsection{Inclusive Decay}
\label{sec:inc}

The inclusive decay $B \to X_s \ell^+\ell^-$ gives access to three independent observables, $H_L$, $H_T$, and $H_A$~\cite{Lee:2006gs}. We will consider the following combinations of these observables:
(a) the branching ratio BR$(B \to X_s \ell^+\ell^-) = H_L + H_T$; 
(b) the normalized lepton forward-backward asymmetry $A_\text{FB} = \frac{3}{4} H_A/(H_L + H_T)$; 
(c) the observable $F_L = H_L/(H_L + H_T)$, in analogy to the longitudinal polarization fraction of the $K^*$ in the $B \to K^* \ell^+ \ell^-$ decay. 
In slight abuse of notation, we denote these observables in the same way in both the $B \to K^* \ell^+ \ell^-$ decay and the $B \to X_s \ell^+ \ell^-$ decay. Analogous to the exclusive decays we will consider a LFU ratio for the branching ratios, and LFU differences for the angular observables $A_\text{FB}$ and $F_L$
\begin{eqnarray}
R_{X_s} &=& \frac{\text{BR}(B \to X_s \mu\mu)}{\text{BR}(B \to X_s ee)} ~, \\
D_{F_L^{X_s}} &=& F_L(B \to X_s \mu\mu) - F_L(B \to X_s ee) ~, \\
D_{A_\text{FB}^{X_s}} &=& A_\text{FB}(B \to X_s \mu\mu) - A_\text{FB}(B \to X_s ee)  ~.
\end{eqnarray}
Expressions for these observables in terms of the Wilson coefficients $\Delta C_9$ and $\Delta C_9^\prime$ are collected in Appendix~\ref{sec:appendix1}.
In contrast to the exclusive decays, the right-handed coefficient $\Delta C_9^\prime$ enters these expressions only at the quadratic level.
The LFU observables in the inclusive decay are therefore complementary to those in the exclusive decays discussed above.

\section{Predictions for Lepton Flavor Universality Violation}
\label{sec:predictions}

\begin{figure*}[th]
\centering
\includegraphics[width=0.5\textwidth]{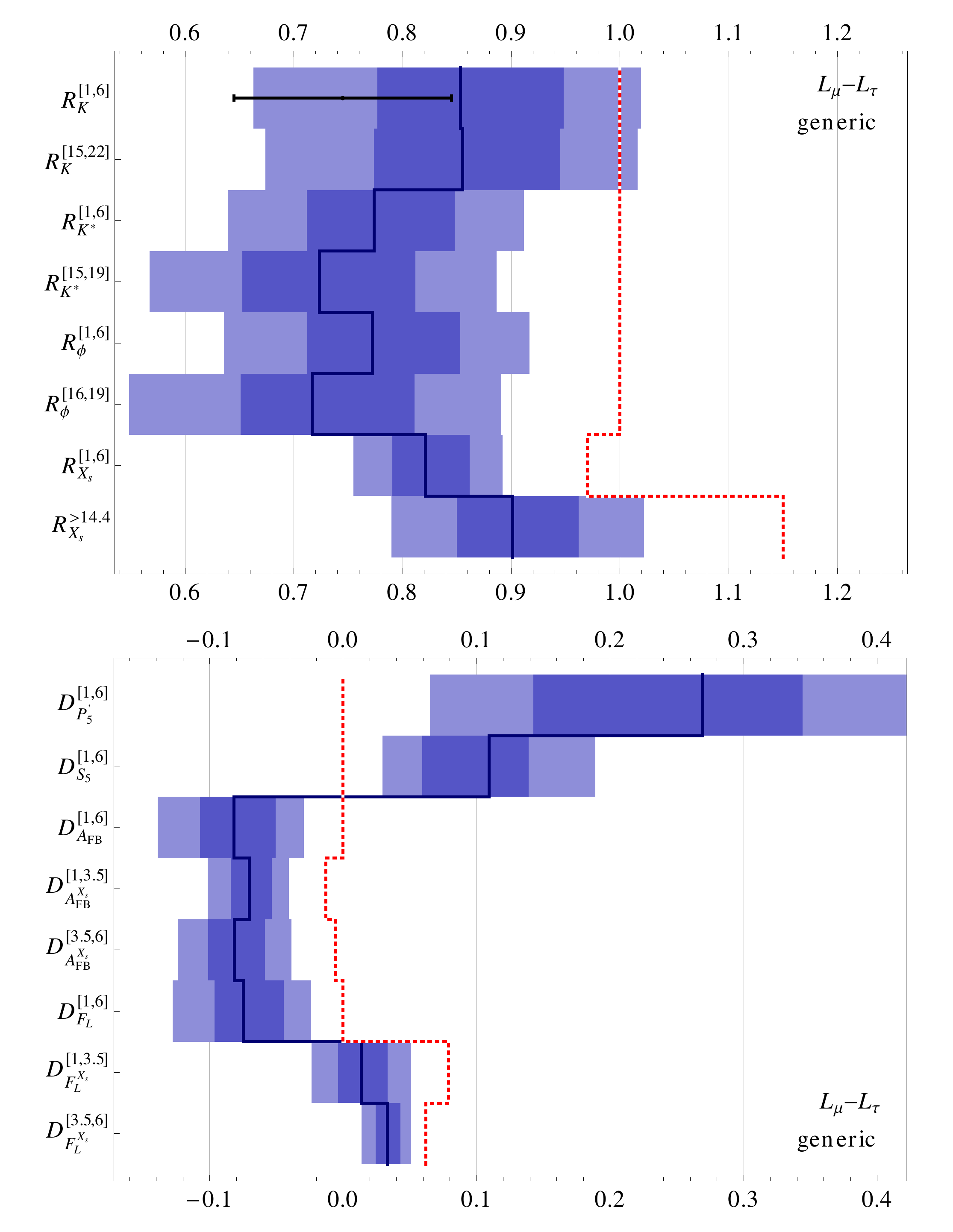}\includegraphics[width=0.5\textwidth]{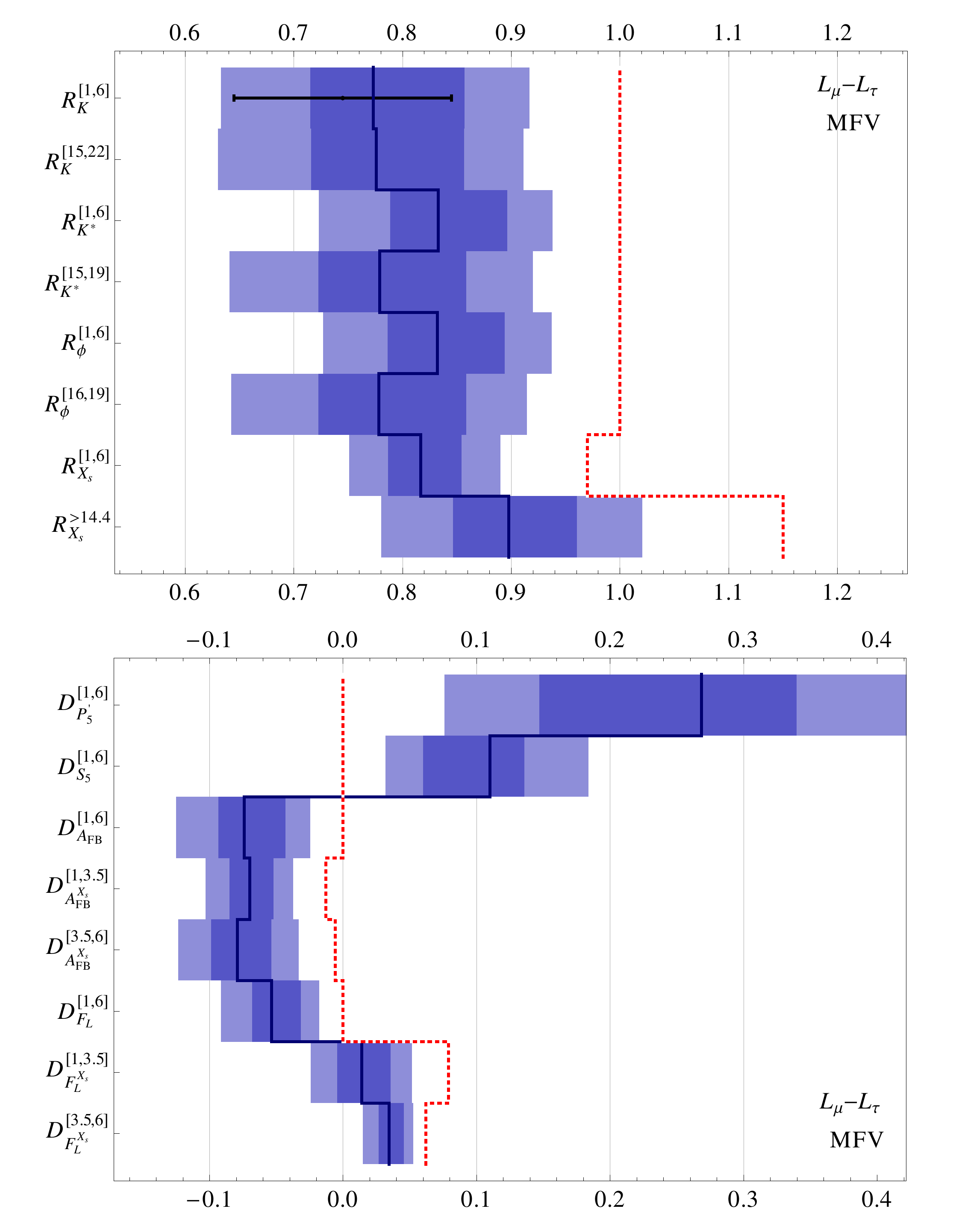}
\caption{Predictions for lepton flavor universality ratios and differences in models with gauged $L_\mu - L_\tau$. Left: generic quark couplings. Right: MFV quark couplings. The solid lines show the predicted central values, the $1\sigma$ and $2\sigma$ ranges are shown by the shaded bands. The SM predictions are indicated by the dotted line. The experimental measurement of $R_K^{[1,6]}$ is also given for comparison.}
\label{fig:LFU}
\end{figure*}

\renewcommand{\arraystretch}{2.4}
\setlength\tabcolsep{0pt}
\begin{table*}[th] 
\begin{center}
\begin{tabularx}{\textwidth}{ccccccc}
\hline\hline
 & \multicolumn{2}{c}{(i) $L_\mu - L_\tau$ generic} & \multicolumn{2}{c}{(ii) $L_\mu - L_\tau$ MFV} & \multicolumn{2}{c}{(iii) LH quarks \& muons}  \\ \hline
 \,~~~~~~~~~~~~~~~~~~~~~~~&~~~~~~~~~~ $1\sigma$ ~~~~~~~~~~&~~~~~~~~~~ $2\sigma$ ~~~~~~~~~~&~~~~~~~~~~ $1\sigma$ ~~~~~~~~~~&~~~~~~~~~~ $2\sigma$ ~~~~~~~~~~&~~~~~~~~~~ $1\sigma$ ~~~~~~~~~~&~~~~~~~~~~ $2\sigma$ \\
\hline\hline
\rowcolor[gray]{.9} $R_K^{[1,6]}$ & $0.85^{+0.09}_{-0.08}$ & $[0.66,1.02]$ & $0.77^{+0.08}_{-0.06}$ & $[0.63,0.92]$ & $0.76^{+0.10}_{-0.09}$ & $[0.55,0.93]$   \\
$R_K^{[15,22]}$ & $0.86^{+0.09}_{-0.08}$ & $[0.67,1.02]$ & $0.78^{+0.08}_{-0.06}$ & $[0.63,0.91]$ & $0.76^{+0.10}_{-0.09}$ & $[0.55,0.93]$   \\
\rowcolor[gray]{.9} $R_{K^*}^{[1,6]}$ & $0.77^{+0.07}_{-0.06}$ & $[0.64,0.91]$ & $0.83^{+0.06}_{-0.04}$ & $[0.72, 0.94]$ & $0.77^{+0.10}_{-0.08}$ & $[0.58,0.93]$   \\
$R_{K^*}^{[15,19]}$ & $0.72^{+0.09}_{-0.07}$ & $[0.57,0.89]$ & $0.78^{+0.08}_{-0.06}$ & $[0.64,0.92]$ & $0.75^{+0.11}_{-0.09}$ & $[0.53,0.92]$ \\
\rowcolor[gray]{.9} $R_{\phi}^{[1,6]}$ & $0.77^{+0.08}_{-0.06}$ & $[0.64,0.92]$ & $0.83^{+0.06}_{-0.05}$ & $[0.73, 0.94]$ & $0.77^{+0.10}_{-0.08}$ & $[0.57,0.93]$   \\
$R_{\phi}^{[16,19]}$ & $0.72^{+0.09}_{-0.07}$ & $[0.55,0.89]$ & $0.78^{+0.08}_{-0.06}$ & $[0.64,0.91]$ & $0.75^{+0.10}_{-0.09}$ & $[0.54,0.93]$   \\
\rowcolor[gray]{.9} $R_{X_s}^{[1,6]}$ & $0.82^{+0.04}_{-0.03}$ & $[0.75,0.89]$ & $0.82^{+0.04}_{-0.03}$ & $[0.75,0.89]$ & $0.73^{+0.08}_{-0.08}$ & $[0.56,0.88]$   \\
$R_{X_s}^{>14.4}$ & $0.90^{+0.06}_{-0.05}$ & $[0.79,1.02]$ & $0.90^{+0.06}_{-0.05}$ & $[0.78,1.02]$ & $0.84^{+0.11}_{-0.11}$ & $[0.62,1.04]$    \\
\hline\hline
\rowcolor[gray]{.9} $D_{P_5^\prime}^{[1,6]}$ & $+0.27^{+0.07}_{-0.13}$ & $[+0.07,+0.47]$ & $+0.27^{+0.07}_{-0.12}$ & $[+0.08,+0.45]$ & $+0.112^{+0.049}_{-0.093}$ & $[-0.014,+0.281]$    \\
$D_{S_5}^{[1,6]}$ & $+0.109^{+0.030}_{-0.050}$ & $[+0.029,+0.189]$ & $+0.110^{+0.026}_{-0.050}$ & $[+0.032,+0.184]$ & $+0.045^{+0.016}_{-0.026}$ & $[+0.008,+0.093]$  \\
\rowcolor[gray]{.9} $D_{A_\text{FB}}^{[1,6]}$ & $-0.082^{+0.031}_{-0.025}$ & $[-0.139,-0.029]$ & $-0.074^{+0.031}_{-0.019}$ & $[-0.125,-0.025]$ & $-0.034^{+0.016}_{-0.015}$ & $[-0.071,-0.007]$   \\
$D_{A_\text{FB}^{X_s}}^{[1,3.5]}$ & $-0.070^{+0.017}_{-0.014}$ & $[-0.102,-0.040]$ & $-0.070^{+0.018}_{-0.015}$ & $[-0.103,-0.037]$ & $-0.043^{+0.014}_{-0.012}$ & $[-0.073,-0.018]$   \\
\rowcolor[gray]{.9} $D_{A_\text{FB}^{X_s}}^{[3.5,6]}$ & $-0.081^{+0.023}_{-0.020}$ & $[-0.124,-0.039]$ & $-0.079^{+0.026}_{-0.020}$ & $[-0.124,-0.033]$ & $-0.035^{+0.013}_{-0.012}$ & $[-0.066,-0.013]$    \\
$D_{F_L}^{[1,6]}$ & $-0.075^{+0.030}_{-0.022}$ & $[-0.128,-0.024]$ & $-0.054^{+0.022}_{-0.015}$ & $[-0.091,-0.018]$ & $-0.017^{+0.010}_{-0.007}$ & $[-0.038,-0.003]$  \\
\rowcolor[gray]{.9} $D_{F_L^{X_s}}^{[1,3.5]}$ & $+0.014^{+0.020}_{-0.017}$ & $[-0.024,+0.051]$ & $+0.014^{+0.022}_{-0.018}$ & $[-0.024,+0.052]$ & $+0.041^{+0.020}_{-0.015}$ & $[-0.001,+0.070]$    \\
$D_{F_L^{X_s}}^{[3.5,6]}$ & $+0.033^{+0.010}_{-0.009}$ & $[+0.014,+0.051]$ & $+0.035^{+0.011}_{-0.008}$ & $[+0.015,+0.053]$ & $+0.065^{+0.003}_{-0.004}$ & $[+0.056,+0.073]$   \\
\hline\hline
\end{tabularx}
\end{center}
\caption{\small Predictions for lepton flavor universality ratios and differences in models with gauged $L_\mu - L_\tau$ and in models with purely left-handed quark and muon currents.}
\label{tab:summary}
\end{table*}

Having defined the observables that probe lepton flavor universality, we go on and provide new physics predictions for these observables. In the context of $L_\mu - L_\tau$, we consider two cases: 
\begin{itemize}
 \item[(i)] the generic case where the mixing Yukawas are arbitrary and both $\Delta C_9$ and $\Delta C_9^\prime$ can be affected significantly by new physics;
 \item[(ii)] the physically motivated case of minimal flavor violation (MFV), where the flavor structure of the left-handed mixing Yukawas are determined by CKM angles, $Y_{Qb} Y_{Qs}^* \propto V_{tb} V_{ts}^*$, and the right-handed mixing Yukawas are negligible. In that case only $C_9$ receives new physics contributions, while $\Delta C_9^\prime \sim 0$.
\end{itemize}   

In~\cite{Altmannshofer:2014rta} the following best fit values for $\Delta C_9^{(\prime)}$ and improvements in the $\chi^2$ were found in a global fit to all relevant $ b \to s \mu^+\mu^- $ data
\begin{align} \label{eq:fit}
 \text{(i)}~ & \Delta C_9 = -1.10 \,,~ \Delta C_9^\prime = 0.45 \,, & \Delta \chi^2 = 15.6 \,, \\
 \text{(ii)}~ & \Delta C_9 = -1.07  \,, & \Delta \chi^2 = 13.7 \,.
\end{align}
We use the corresponding best fit ranges for $\Delta C_9^{(\prime)}$ determined in~\cite{Altmannshofer:2014rta} to make predictions for the LFU observables defined in Sec.~\ref{sec:LFU}.

The results are summarized in Fig.~\ref{fig:LFU} and Tab.~\ref{tab:summary}.
Shown are the predicted central values together with the $1\sigma$ and $2\sigma$ ranges. In Fig.~\ref{fig:LFU} also the SM predictions and the experimental result for $R_K^{[1,6]}$ are shown for comparison.
For completeness we also include predictions for the lepton flavor difference of the angular observable $P_5^\prime$
\begin{equation}
 D_{P_5^\prime} = P_5^\prime(B \to K^* \mu\mu) - P_5^\prime(B \to K^* ee) ~,
\end{equation}
with $P_5^\prime = S_5 /\sqrt{F_L(1-F_L)}$. 

We observe that both the LFU ratios, $R_i$, and the LFU differences, $D_i$, differ significantly from the respective SM predictions. 
All branching ratio ratios that we consider are below 1, with central values ranging approximately between $\sim 0.75$ and  $\sim 0.9$.
The predictions for $D_{P_5^\prime}$ and $D_{S_5}$ are above their SM values, while the various differences of $A_\text{FB}$ and $F_L$ are predicted below their SM values.
We expect that within the next few years of LHCb data taking the experimental sensitivity should be high enough to either confirm or exclude the predicted non-standard values for $R_K$ and $R_{K^*}$.
Given that $B_s$-meson production is suppressed compared to $B$-meson production at the LHC, establishing a deviation of $R_\phi$ from 1 will be more challenging. 
At Belle II, the predicted non-standard values for $R_{X_s}$ should be well within reach. 
Concerning the LFU differences, it has to be seen if experimental sensitivities at LHCb and Belle II will be sufficient to observe also the predicted deviations in $D_{P_5^\prime}$, $D_{S_5}$, $D_{A_\text{FB}}$, and $D_{F_L}$ in the near future.

We now comment on the differences between the generic case (i) and the MFV case (ii).
Notable differences do exist in the predictions for the ratios $R_K$, $R_{K^*}$ and $R_\phi$.
Because of the parity of the final state mesons, the ratios are corrected by the combination $C_9+C_9^\prime$ ($R_K$) or mainly by $C_9-C_9^\prime$ ($R_{K^*}$ and $R_\phi$). As the global fit results~(\ref{eq:fit}) show preference for a negative value for $C_9$ and a slightly positive value for $C_9^\prime$, this leads to the MFV prediction to be consistently higher (lower) than the generic prediction for $ R_{K^*}$ and $R_{\phi}$ ($R_K$).
As the inclusive decay $B \to X_s \ell^+\ell^-$ depends on the $\Delta C_9^\prime$ only at the quadratic level, the predictions for the corresponding LFU observables are virtually identical in the case (i) and case (ii).
A $\sim 20\%$ suppression of $R_{X_s}$ compared to the SM prediction is therefore a very robust prediction of the $L_\mu - L_\tau$ scenario.
We observe that also the predictions for the $B \to K^* \ell^+\ell^-$ angular differences $D_{P_5^\prime}$, $D_{S_5}$, and $D_{A_\text{FB}}$, are very similar in the cases (i) and (ii).

\begin{figure}[th]
\centering
\includegraphics[width=0.49\textwidth]{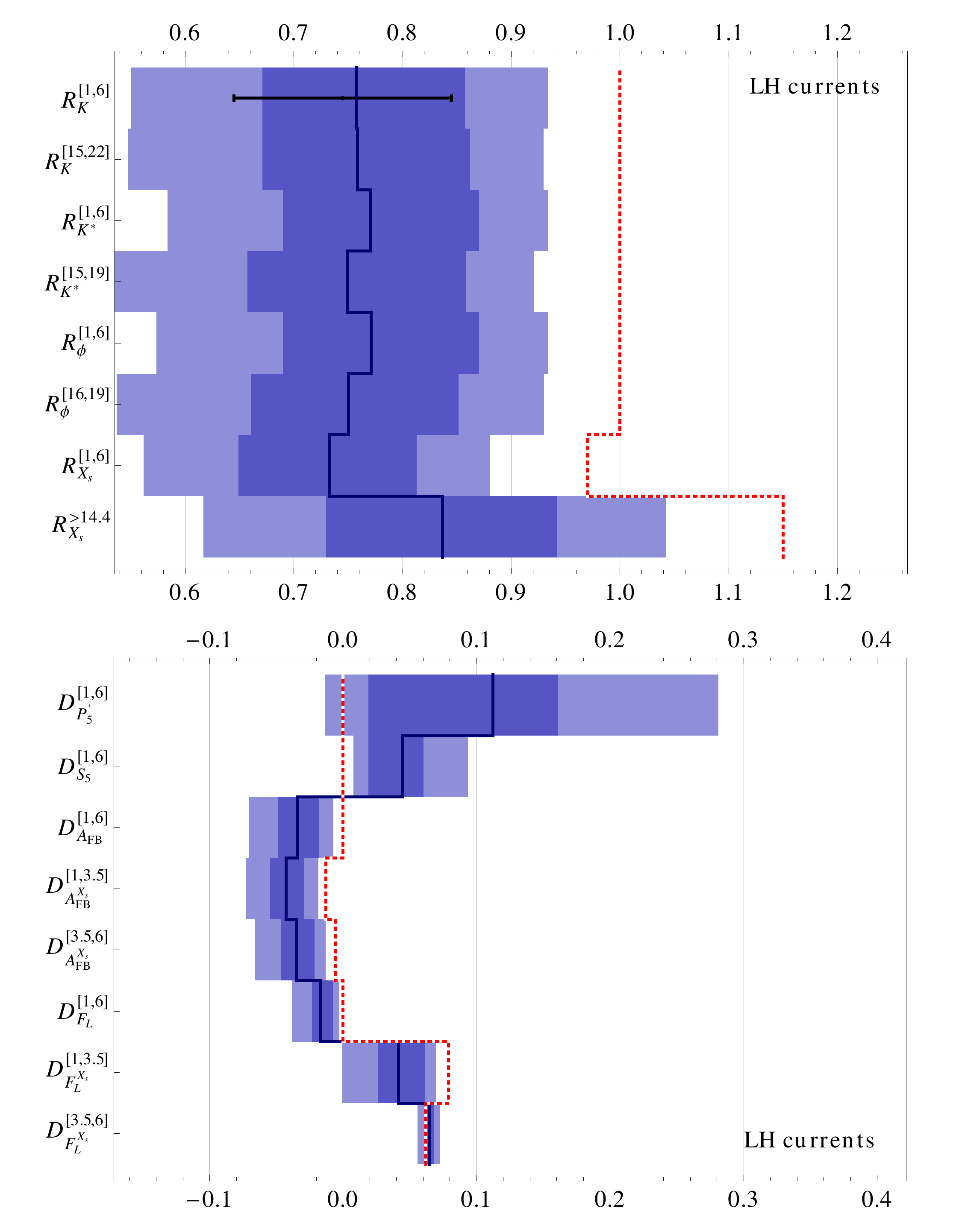}
\caption{Predictions for lepton flavor universality ratios and differences in models with purely left-handed quark and muon currents. The solid lines show the predicted central values, the $1\sigma$ and $2\sigma$ ranges are shown by the shaded bands. The SM predictions are indicated by the dotted line. The experimental measurement of $R_K^{[1,6]}$ is also given for comparison.}
\label{fig:LFU_LH}
\end{figure}

Aside from comparison with the SM predictions, the lepton flavor universality observables can also be useful in distinguishing between alternative models of new physics.
Here, we investigate to what extent the $L_\mu - L_\tau$ setup can be distinguished from a scenario where the anomalies in $B \to K^* \mu^+\mu^-$ are resolved by NP in an operator that consists of left-handed quark and muon currents
\begin{equation}
 \mathscr{H}_\text{eff} =  - \frac{4G_F}{\sqrt{2}} V_{tb} V_{ts}^* \frac{e^2}{16\pi^2} \Delta C_L (\bar s \gamma_\alpha P_L b) (\bar \mu \gamma^\alpha P_L \mu) ~.
\end{equation}
Various NP scenarios of this type have been discussed recently~\cite{Glashow:2014iga,Bhattacharya:2014wla,Niehoff:2015bfa,Greljo:2015mma}.
Indeed, compared to scenarios with NP in $\Delta C_9$, NP in $\Delta C_L$ gives only slightly worse fits to the data~\cite{Altmannshofer:2014rta}
\begin{align} \label{eq:fitLH}
 \Delta C_L = -1.06  ~,~~ \Delta \chi^2 = 9.8 ~.
\end{align}
Expressions for the LFU observables in such a scenario are given in Appendix~\ref{sec:appendix2}.

In Fig.~\ref{fig:LFU_LH} we summarize the predicted ranges for the LFU ratios and LFU differences in this setup.
The results are also collected in the last columns of Tab.~\ref{tab:summary}.
The ranges for the LFU ratios, $R_i$, almost entirely include the ranges found in the cases (i) and (ii) discussed above. This is due to the larger range for $\Delta C_L$ that is allowed by the global fit~\cite{Altmannshofer:2014rta}.
We observe that the non-standard effects in the LFU differences, $D_i$, are considerably smaller compared to the cases (i) and (ii). Precision measurements of these observables might allow to distinguish the $L_\mu - L_\tau$ scenarios from NP purely in left-handed currents.

\section{Summary} \label{sec:sum}

In this short paper we provided explicit expressions and numerical predictions for a variety of lepton flavor universality observables associated with the model of Ref.~\cite{Altmannshofer:2014cfa} based on gauged $L_\mu - L_\tau$. We considered known LFU ratios of branching ratios as well as LFU differences of angular observables. 

Generically, a robust prediction is that all branching ratio ratios are suppressed compared to the SM prediction by $O(20\%)$, both in the case of generic quark couplings and in the case of quark couplings that follow the principle of minimal flavor violation. The predictions for the $B \to K \ell^+\ell^-$ decay, $R_K =0.85^{+0.09}_{-0.08}$ (generic quark couplings) and $R_K = 0.77^{+0.08}_{-0.06}$ (MFV quark couplings), have recently received experimental support from the measurement of this ratio by LHCb~\cite{Aaij:2014ora}. The LFU ratios represent a very clear target of opportunity for future measurements in the search for new physics. 

The LFU differences of angular observables provide additional means to probe new physics in rare $B$ decays. We predict non-standard effects at the level of $O(10\%)$ in the context of gauged $L_\mu - L_\tau$.
These observables might also serve as discriminants between alternative new physics models if precision measurements will become available in the future.

\begin{acknowledgments}
Research at Perimeter Institute is supported by the Government of Canada through Industry Canada and by the Province of Ontario through the Ministry of Research and Innovation. IY is supported in part by funds from the NSERC of Canada. 
\end{acknowledgments}

\begin{appendix}\onecolumngrid
\section{Numerical Expressions in the \texorpdfstring{\boldmath $L_\mu - L_\tau$}{Lmu - Ltau} Scenario} \label{sec:appendix1}

For the branching ratio ratios of the exlcusive decays we find
\begin{eqnarray} \label{eq:LFU1}
 R_K^{[1,6]} &\simeq& 1.00 + 0.243 (\Delta C_9 + \Delta C_9^\prime) + 0.030 (\Delta C_9 + \Delta C_9^\prime)^2 ~, \\
 R_K^{[15,22]} &\simeq& 1.00 + 0.241 (\Delta C_9 + \Delta C_9^\prime) + 0.030 (\Delta C_9 + \Delta C_9^\prime)^2 ~, \\[10pt]
 R_{K^*}^{[1,6]} &\simeq& 1.00 + 0.192 \Delta C_9 - 0.198 \Delta C_9^\prime + 0.034 \left( (\Delta C_9)^2 + (\Delta C_9^\prime)^2 \right) - 0.052 \Delta C_9 \Delta C_9^\prime~, \\
 R_{K^*}^{[15,19]} &\simeq& 1.00 + 0.241 \Delta C_9 - 0.182 \Delta C_9^\prime + 0.033 \left( (\Delta C_9)^2 + (\Delta C_9^\prime)^2 \right) - 0.049 \Delta C_9 \Delta C_9^\prime~, \\[10pt]
 R_{\phi}^{[1,6]} &\simeq& 1.00 + 0.193 \Delta C_9 - 0.200 \Delta C_9^\prime + 0.034 \left( (\Delta C_9)^2 + (\Delta C_9^\prime)^2 \right) - 0.052 \Delta C_9 \Delta C_9^\prime~, \\
 R_{\phi}^{[16,19]} &\simeq& 1.00 + 0.241 \Delta C_9 - 0.201 \Delta C_9^\prime +0.032 \left( (\Delta C_9)^2 + (\Delta C_9^\prime)^2 \right) - 0.053 \Delta C_9 \Delta C_9^\prime ~, \label{eq:LFU6}
\end{eqnarray}
in reasonable agreement with~\cite{Hiller:2014ula}.
The expressions hold for real Wilson coefficients $\Delta C_9$ and $\Delta C_9^\prime$. A study of CP violating scenarios is beyond the scope of the present work.
The theoretical uncertainty on the leading piece (unity) is expected to be at the sub-percent level, whereas we estimate that the uncertainty on the remaining numerical coefficients in Eqs.~(\ref{eq:LFU1})-(\ref{eq:LFU6}) is at the $10\%$ level. In our numerical analysis in Sec.~\ref{sec:predictions} we neglect the uncertainty of the leading piece and use uncorrelated $20\%$ relative uncertainties for the remaining numerical coefficients.
These uncertainties have only minor impact on our results, as the dominant uncertainty arises from the allowed ranges of  $\Delta C_9$ and $\Delta C_9^\prime$.

For the differences of angular observables in $B \to K^* \ell^+\ell^-$ we obtain
\begin{eqnarray} \label{eq:LFU7}
 D_{A_\text{FB}}^{[1,6]} &\simeq& \frac{0.00 + 0.057 \Delta C_9 }{ 1 + 0.192 \Delta C_9 - 0.198 \Delta C_9^\prime + 0.034 \left( (\Delta C_9)^2 + (\Delta C_9^\prime)^2 \right) - 0.052 \Delta C_9 \Delta C_9^\prime}  ~, \\[10pt]
 D_{F_L}^{[1,6]} &\simeq& \frac{0.00 + 0.039  \Delta C_9 - 0.035 \Delta C_9^\prime - 0.002 \left( (\Delta C_9)^2 + (\Delta C_9^\prime)^2 \right) - 0.008 \Delta C_9 \Delta C_9^\prime}{ 1 + 0.192 \Delta C_9 - 0.198 \Delta C_9^\prime + 0.034 \left( (\Delta C_9)^2 + (\Delta C_9^\prime)^2 \right) - 0.052 \Delta C_9 \Delta C_9^\prime} ~, \\[10pt]
 D_{S_5}^{[1,6]} &\simeq& \frac{0.00 - 0.079 \Delta C_9 - 0.033 \Delta C_9^\prime + 0.006 \left( (\Delta C_9)^2 + (\Delta C_9^\prime)^2 \right)-0.009 \Delta C_9 \Delta C_9^\prime}{ 1 + 0.192 \Delta C_9 - 0.198 \Delta C_9^\prime + 0.034 \left( (\Delta C_9)^2 + (\Delta C_9^\prime)^2 \right) - 0.052 \Delta C_9 \Delta C_9^\prime} ~, \label{eq:LFU9}
\end{eqnarray}
valid for real Wilson coefficients.
Similar to the case of the LFU ratios in~(\ref{eq:LFU1})-(\ref{eq:LFU6}), the leading piece (zero) can be predicted with very high accuracy. The uncertainty on the remaining numerical coefficients is expected to be at the $10\%$ level. In the numerical analysis, we use uncorrelated $20\%$ relative uncertainties for the corresponding numerical coefficients in~(\ref{eq:LFU7})-(\ref{eq:LFU9}).

For the LFU observables in the inclusive decay we make use of the expressions in~\cite{Huber:2015sra,Huber:2005ig} and arrive at
\begin{eqnarray} \label{eq:LFU10}
 R_{X_s}^{[1,6]} &\simeq& (R_{X_s}^{[1,6]})_\text{SM} + 0.181 \Delta C_9 + 0.036 \left( (\Delta C_9)^2 + (\Delta C_9^\prime)^2 \right)~, \\
 R_{X_s}^{>14.4} &\simeq& (R_{X_s}^{>14.4})_\text{SM} + 0.281 \Delta C_9 + 0.042 \left( (\Delta C_9)^2 + (\Delta C_9^\prime)^2 \right) ~, \\[10pt]
 D_{F_L^{X_s}}^{[1,3.5]} &\simeq& \frac{(D_{F_L}^{[1,3.5]})_\text{SM} + 0.065 \Delta C_9 + 0.003 \left( (\Delta C_9)^2 + (\Delta C_9^\prime)^2 \right)}{1 + 0.167 \Delta C_9 + 0.035 \left( (\Delta C_9)^2 + (\Delta C_9^\prime)^2 \right)} ~, \\
 D_{F_L^{X_s}}^{[3.5,6]} &\simeq& \frac{(D_{F_L}^{[3.5,6]})_\text{SM} + 0.031 \Delta C_9 - 0.001 \left( (\Delta C_9)^2 + (\Delta C_9^\prime)^2 \right)}{1 + 0.212 \Delta C_9 + 0.039 \left( (\Delta C_9)^2 + (\Delta C_9^\prime)^2 \right)} ~, \\[10pt]
 D_{A_\text{FB}^{X_s}}^{[1,3.5]} &\simeq& \frac{(D_{A_\text{FB}}^{[1,3.5]})_\text{SM} + 0.050 \Delta C_9 + 0.003 \left( (\Delta C_9)^2 + (\Delta C_9^\prime)^2 \right) }{1 + 0.167 \Delta C_9 + 0.035 \left( (\Delta C_9)^2 + (\Delta C_9^\prime)^2 \right)}~, \\
 D_{A_\text{FB}^{X_s}}^{[3.5,6]} &\simeq& \frac{(D_{A_\text{FB}}^{[3.5,6]})_\text{SM} + 0.052 \Delta C_9 - 0.003 \left( (\Delta C_9)^2 + (\Delta C_9^\prime)^2 \right) }{1 + 0.212 \Delta C_9 + 0.039 \left( (\Delta C_9)^2 + (\Delta C_9^\prime)^2 \right)}~, \label{eq:LFU16}
\end{eqnarray}
valid for real Wilson coefficients.
For the SM predictions we find from~\cite{Huber:2015sra}
\begin{eqnarray}\label{eq:SM1}
(R_{X_s}^{[1,6]})_\text{SM} = 0.970 ~&,&~ (R_{X_s}^{>14.4})_\text{SM} = 1.151~,\\
(D_{F_L^{X_s}}^{[1,3.5]})_\text{SM} = 0.079 ~&,&~ (D_{F_L^{X_s}}^{[3.5,6]})_\text{SM} = 0.062 ~, \\
(D_{A_\text{FB}^{X_s}}^{[1,3.5]})_\text{SM} = -0.010 ~&,&~ (D_{A_\text{FB}^{X_s}}^{[3.5,6]})_\text{SM} = -0.005 ~.\label{eq:SM3}
\end{eqnarray}
We expect uncertainties of $O(0.01)$ for these SM predictions and we neglect them in our numerical analysis.
The small differences in the SM predictions of $R_{X_s}$ from unity and of $D_{F_L}$ and $D_{A_\text{FB}}$ from zero, arise from logarithmically enhanced QED corrections. 
Given the good theoretical control in the prediction of the inclusive $B \to X_s \ell^+\ell^-$ decay~\cite{Huber:2015sra}, the uncertainties on the remaining numerical coefficients in Eqs.~(\ref{eq:LFU10})-(\ref{eq:LFU16}) are likely below $10\%$. We use uncorrelated $10\%$ relative uncertainties in our numerical analysis. This choice of uncertainties has only very minor impact on our results.

\section{Numerical Expressions in the Scenario with Left-Handed Currents} \label{sec:appendix2}

In the scenario with NP in left-handed currents only, we find for the LFU observables in exclusive decays
\begin{eqnarray} 
 R_K^{[1,6]} &\simeq& 1.00 + 0.245 \Delta C_L + 0.015 (\Delta C_L)^2 ~, \\
 R_K^{[15,22]} &\simeq& 1.00 + 0.244 \Delta C_L + 0.015 (\Delta C_L)^2 ~, \\[10pt]
 R_{K^*}^{[1,6]} &\simeq& 1.00 + 0.235 \Delta C_L + 0.017 (\Delta C_L)^2 ~, \\
 R_{K^*}^{[15,19]} &\simeq& 1.00 + 0.254 \Delta C_L + 0.016 (\Delta C_L)^2~, \\[10pt]
 R_{\phi}^{[1,6]} &\simeq& 1.00 + 0.234 \Delta C_L + 0.017 (\Delta C_L)^2~, \\
 R_{\phi}^{[16,19]} &\simeq& 1.00 + 0.253 \Delta C_L + 0.016 (\Delta C_L)^2~, 
\end{eqnarray}
\begin{eqnarray} 
 D_{A_\text{FB}}^{[1,6]} &\simeq& \frac{0.00 + 0.029 \Delta C_L + 0.003 (\Delta C_L)^2}{ 1 + 0.235 \Delta C_L  + 0.017 (\Delta C_L)^2}  ~, \\[10pt]
 D_{F_L}^{[1,6]} &\simeq& \frac{0.00 + 0.011 \Delta C_L - 0.001 (\Delta C_L)^2}{ 1 + 0.235 \Delta C_L  + 0.017 (\Delta C_L)^2}  ~, \\[10pt]
 D_{S_5}^{[1,6]} &\simeq& \frac{0.00 - 0.037 \Delta C_L -0.004 (\Delta C_L)^2}{ 1 + 0.235 \Delta C_L  + 0.017 (\Delta C_L)^2}  ~. 
\end{eqnarray}
Our treatment of uncertainties in the numerical analysis is equivalent to the $L_\mu - L_\tau$ scenario discussed above.

For the LFU observables in the inclusive decay we find from~\cite{Huber:2015sra,Huber:2005ig}
\begin{eqnarray} 
 R_{X_s}^{[1,6]} &\simeq& (R_{X_s}^{[1,6]})_\text{SM} + 0.243 \Delta C_L + 0.018 (\Delta C_L)^2 ~, \\
 R_{X_s}^{>14.4} &\simeq& (R_{X_s}^{>14.4})_\text{SM} + 0.319 \Delta C_L + 0.022 (\Delta C_L)^2 ~, \\[10pt]
 D_{F_L^{X_s}}^{[1,3.5]} &\simeq& \frac{(D_{F_L}^{[1,3.5]})_\text{SM} + 0.046 \Delta C_L + 0.002 (\Delta C_L)^2}{1 + 0.236 \Delta C_L + 0.018 (\Delta C_L)^2} ~, \\
 D_{F_L^{X_s}}^{[3.5,6]} &\simeq& \frac{(D_{F_L}^{[3.5,6]})_\text{SM} + 0.014 \Delta C_L - 0.0002 (\Delta C_L)^2}{1 + 0.272 \Delta C_L + 0.020 (\Delta C_L)^2} ~, \\[10pt]
 D_{A_\text{FB}^{X_s}}^{[1,3.5]} &\simeq& \frac{(D_{A_\text{FB}}^{[1,3.5]})_\text{SM} + 0.026 \Delta C_L + 0.004 (\Delta C_L)^2}{1 + 0.236 \Delta C_L + 0.018 (\Delta C_L)^2}~, \\
 D_{A_\text{FB}^{X_s}}^{[3.5,6]} &\simeq& \frac{(D_{A_\text{FB}}^{[3.5,6]})_\text{SM} + 0.022 \Delta C_L + 0.003 (\Delta C_L)^2 }{1 + 0.272 \Delta C_L + 0.020 (\Delta C_L)^2}~.
\end{eqnarray}
The SM predictions have already been given above in Eqs.~(\ref{eq:SM1}) - (\ref{eq:SM3}), and our treatment of uncertainties in the numerical analysis is equivalent to the $L_\mu - L_\tau$ scenario.

\end{appendix}
\twocolumngrid



\end{document}